\pdfoutput=1

\documentclass[10pt,prd,aps,amssymb,amsmath,tightenlines,showkeys,twocolumn]{revtex4}

\usepackage{graphicx}
\usepackage[T1]{fontenc}
\usepackage[latin1]{inputenc}
\usepackage[english]{babel}
\usepackage{amsmath}
\usepackage{amsfonts}
\usepackage{amssymb}
\usepackage{dsfont}
\usepackage{color}
\usepackage{slashed}
\usepackage{bbm}
\usepackage{bbold}

\def\cP{\mathcal P}
\def\cT{\mathcal T}

\def\cPT{\mathcal{PT}}
\def\cL{\mathcal{L}}

\def\half{{\textstyle{\frac{1}{2}}}}
\def\one{\mathds 1}
\def\vep{\varepsilon}

\begin{document}
\title{Relativistic $\cPT$-symmetric fermionic theories in 1+1 and 3+1
dimensions}

\author{Alireza Beygi}\email{beygi@thphys.uni-heidelberg.de}
\author{S. P. Klevansky}\email{spk@physik.uni-heidelberg.de}
\affiliation{Institut f\"ur Theoretische Physik, Universit\"at Heidelberg,
Philosophenweg 12, 69120 Heidelberg, Germany}

\author{Carl M. Bender}\email{cmb@wustl.edu}
\affiliation{Department of Physics, Washington University, St. Louis,
Missouri 63130, USA}

\begin{abstract}
Relativistic $\cPT$-symmetric fermionic interacting systems are studied in 1+1
and 3+1 dimensions. The noninteracting Dirac equation is separately $\cP$ and
$\cT$ invariant. The objective here is to include non-Hermitian $\cPT$-symmetric
interaction terms that give rise to {\it real} spectra. Such interacting systems
could be physically realistic and could describe new physics. The simplest such
non-Hermitian Lagrangian density is $\cL=\cL_0+\cL_{\rm int}=\bar\psi(i\slashed
\partial-m)\psi-g\bar\psi\gamma^5\psi$. The associated relativistic Dirac
equation is $\cPT$ invariant in 1+1 dimensions and the associated Hamiltonian
commutes with $\cPT$. However, the dispersion relation $p^2=m^2-g^2$ shows that
the $\cPT$ symmetry is broken (the eigenvalues become complex) in the chiral
limit $m\to0$. For field-theoretic interactions of the form $\cL_{\rm int}=-g(
\bar\psi\gamma^5\psi)^N$ with $N=2,\,3$, which we can only solve approximately,
we also find that if the associated (approximate) Dirac equation is $\cPT$
invariant, the dispersion relation always gives rise to complex energies in the
chiral limit $m\to0$. Other models are studied in which $x$-dependent
$\cPT$-symmetric potentials such as $ix^3$, $-x^4$, $i\kappa/x$, Hulth{\'e}n, or
periodic potentials are coupled to the fermionic field $\psi$ using vector or
scalar coupling schemes or combinations of both. For each of these models the
classical trajectories in the complex-$x$ plane are examined. Some combinations
of these potentials can be solved numerically, and it is shown explicitly that a
real spectrum can be obtained. In 3+1 dimensions, while the simplest system $\cL
=\cL_0+\cL_{\rm int}=\bar\psi(i\slashed\partial-m)\psi-g\bar\psi\gamma^5\psi$
resembles the 1+1-dimensional case, the Dirac equation is {\it not} $\cPT$
invariant because $\cT^2=-\one$. This explains the appearance of complex
eigenvalues as $m\to0$. Other Lorentz-invariant two-point and four-point
interactions are considered that give non-Hermitian $\cPT$-symmetric terms in
the Dirac equation. Only the axial vector and tensor Lagrangian interactions
$\cL_{\rm int}=-i\bar\psi\tilde B_\mu\gamma^5\gamma^\mu\psi$ and $\cL_{\rm int}=-i\bar
\psi T_{\mu\nu}\sigma^{\mu\nu}\psi$ fulfil both requirements of $\cPT$
invariance of the associated Dirac equation and non-Hermiticity. The dispersion
relations show that both interactions lead to complex spectra in the chiral
limit $m\to0$. The effect on the spectrum of the additional constraint of
selfadjointness of the Hamiltonian with respect to the $\cPT$ inner product
is investigated.
\end{abstract}

\keywords{$\cPT$ symmetry, relativistic fermionic theories, Dirac equation}
\maketitle

\section{Introduction}\label{s1}
A non-Hermitian quantum-mechanical Hamiltonian $H$ that is invariant under
combined parity (space reflection) $\cP$ and time reversal $\cT$ can have real
eigenvalues \cite{r1,r1a}. If the spectrum is entirely real, we say that $H$ has
an {\it unbroken} $\cPT$ symmetry. However, if $H$ has complex eigenvalues, we
say that $H$ has a {\it broken} $\cPT$ symmetry. Numerous theoretical studies of
classical and quantum-mechanical $\cPT$-symmetric systems have been done and
many experiments on such systems have been performed. The remarkable features of
$\cPT$-symmetric include $\cPT$ symmetry breaking in coupled wave guides,
unidirectional invisibility, enhanced sensing at exceptional points, level
bifurcation in superconducting wires, and robust wireless power transfer
\cite{r2,r3,r4,r5,r6,r7,r8,r9}.

In quantum mechanics $x\to-x$ under parity $\cP$ and $i\to-i$ under time
reversal $\cT$. Thus, the quantum-mechanical Hamiltonian $H=p^2+x^2(ix^\vep)$
($\vep$ real) is $\cPT$ invariant; $H$ has a {\it real positive discrete}
spectrum when $\vep\geq0$ \cite{r1}. This quantum theory generalizes to
relativistic quantum field theory if the operator $x(t)$ is replaced by the
pseudoscalar field $\phi(t,{\bf x})$ so that $\phi(t,{\bf x})\to-\phi(t,{-\bf x}
)$ under $\cP$ and $\phi(t,{\bf x})\to\phi^*(-t,{\bf x})$ under $\cT$. The
analogous bosonic field-theoretic Hamiltonian density $(\partial\phi)^2+\phi^2(
i\phi)^\vep$ also appears to have a real spectrum; this was shown to first order
in $\vep$ for $0\leq D<2$ \cite{r10}.

While $\cPT$-symmetric bosonic systems have been studied heavily (there are over
4,000 papers on such systems), only a few papers have been written on
$\cPT$-symmetric fermionic systems. Early work on matrix models of fermionic
systems can be found in Refs.~\cite{r11,r13,r12,r14}. The Lagrangian density for
a free relativistic fermionic field with mass $m$ was extended by including a
non-Hermitian axial mass term $\cL_{\rm int}=-g\bar\psi\gamma^5\psi$, where $g$
is a real mass parameter \cite{r15}. Further developments were made in
Ref.~\cite{r16} in which quantum electrodynamics was extended to include such a
term and the restoration of gauge symmetry was investigated. In Ref.~\cite{R14}
the relationship between conserved currents and invariances of the Lagrangian in
the framework of non-Hermitian field theories was examined. An application of
$\cPT$-symmetric fermionic field theory to neutrino species oscillation was
proposed in Ref.~\cite{R15} in which an 8-dimensional Dirac equation was
analyzed. Neutrino oscillations in the context of $\cPT$ symmetry were studied
further in Ref.~\cite{R16}.

$\cPT$-symmetric fermionic field theories in 1+1 dimensions share the property
with quantum-mechanical and bosonic field theories that $\cT^2=\one$ \cite{r15}.
However, in Ref.~\cite{r12} it was noted that $\cPT$-symmetric fermionic systems
in 3+1 dimensions have the propery that $\cT^2=-\one$. To explain this we first
examine what happens in 1+1 dimensions, where the gamma matrices are
\cite{r19}
\begin{equation}
\gamma^0=\left(\begin{array}{cc}0 & 1\\ 1 & 0\\ \end{array}\right),\qquad
\gamma^1=\left(\begin{array}{cc} 0 & 1\\ -1 & 0\\ \end{array}\right).
\label{E1}
\end{equation}
Note that $\big(\gamma^0\big)^2=\one$, $\big(\gamma^1\big)^2=-\one$, and
$\gamma^5=\gamma^0\gamma^1=-\sigma_3$, where $\sigma_3$ is a Pauli matrix.
Let us identify the discrete spatial symmetries of the free Dirac equation
\begin{equation}
\big[i\gamma^0\partial_0+i\gamma^1\partial_1-m\big]\psi(t,x)=0.
\label{E2}
\end{equation}
(Here $\partial_0=\partial_t$ and $\partial_1=\partial_x$.) To determine the
effect of a space reflection we let $x\to-x$ and then multiply (\ref{E2}) on the
left by $\gamma^0$ to get
$$\big[i\gamma^0\partial_0+i\gamma^1\partial_1-m\big]\gamma^0\psi(t,-x)=0.$$
Because this equation has the {\it same form} as (\ref{E2}) we identify that the
action of parity reflection $\cP$ on the spinor $\psi(t,x)$ is given by
\begin{equation}
\cP:\,\psi(t,x)\to\cP\psi(t,x)\cP^{-1}=\gamma^0\psi(t,-x).
\label{E2a}
\end{equation}

Next, to determine the effect of time reversal $\cT$ we let $t\to-t$ in
(\ref{E2}), take the complex conjugate of the resulting equation, and again
multiply on the left by $\gamma^0$. We get
$$\big[i\gamma^0\partial_0+i\gamma^1\partial_1-m\big]\gamma^0\psi^*(-t,x)=0.$$
Again, from form invariance we conclude that time reversal for spinors in 1+1
dimensions is given by
\begin{equation}
\cT:\,\psi(t,x)\to\cT\psi(t,x)\cT^{-1}=\gamma^0\psi^*(-t,x).
\label{E2b}
\end{equation}
Since $\gamma^0$ is real we see that applying $\cP$ or $\cT$ twice leaves
$\psi(t,x)$ invariant. Thus, $\cP^2=\one$ and $\cT^2=\one$. (Interestingly, this
property of time reversal in 1+1 dimensions implies that the Dirac electron
behaves like a boson \cite{thaller}.)

In 3+1 dimensions the Dirac representation of the gamma matrices is \cite{R18}
\begin{eqnarray}
\gamma^0&=&\left(\begin{array}{cc}\one& 0\\ 0 & -\one\end{array}\right),\qquad
\gamma^i=\left(\begin{array}{cc} 0& \sigma^i\\ -\sigma^i &0\end{array}\right),
\nonumber\\
&&\quad\gamma^5=i\gamma^0\gamma^1\gamma^2\gamma^3=\left(\begin{array}{cc} 0&
\one\\ \one & 0\end{array}\right),
\label{E3}
\end{eqnarray}
where $\sigma^i$ are the Pauli matrices. The actions of parity and time
reversal obtained similarly, are now \cite{R18}
\begin{eqnarray}
\cP:\,\psi(t,{\bf x}) &\to& \cP\psi(t,{\bf x})\cP^{-1}=\gamma^0\psi(t,-{\bf x}),\nonumber\\
\cT:\,\psi(t,{\bf x}) &\to& \cT\psi(t,{\bf x})\cT^{-1}=i\gamma^1\gamma^3\psi^*(-t,{\bf x}).
\label{E4}
\end{eqnarray}
If we apply $\cT$ twice, we observe a change of sign: $\cP^2=\one$, but now
$\cT^2=-\one$. This underscores the different nature of fermions in 3+1
dimensions.

The purpose of this paper is to investigate the behavior of 1+1- and
3+1-dimensional relativistic $\cPT$-invariant fermionic theories. An exploratory
study in Ref.~\cite{R19} examined in part the properties of a $\cPT$-symmetric
fermionic Lee model. This paper begins by reexamining the results in \cite{r15},
where it was assumed that for real $g$ the Lagrangian $\cL=\bar\psi(i\slashed
\partial-m-g\gamma^5)\psi$ is $\cPT$ symmetric. We find that including the
axial term gives a dispersion relation $p^2=m^2-g^2$ that yields a real value
for the physical mass only when $m^2\geq g^2$. This implies that the spectrum is
not real in the chiral limit $m\to0$. This result holds for Lagrangians in both
1+1 and 3+1 dimensions. We ask, Why is this so and under what conditions is it
not so? Obtaining spectral reality in the chiral limit is part of the motivation
for this paper. One of our long-range goals to construct a $\cPT$-symmetric
version of the Thirring and Nambu-Jona-Lasinio models. The challenge is to
identify additional non-Hermitian terms that are both $\cPT$ symmetric and
chiral and give rise to a real spectrum in the chiral limit \cite{R20}.

Two ingredients are required for a precise analysis of fermionic systems: (i)
Care must be taken in analyzing time reversal, which is nontrivial for fermionic
systems; (ii) care is needed in deciding on the form of $\cPT$-adjoint
operators. In this paper we focus first on time reversal and then address the
constraint of selfadjointness with respect to the $\cPT$ inner product for
fermions.

For various interactions in 1+1 and 3+1 dimensions we use the Euler-Lagrange
equations to construct the Dirac equation that results from a Lagrangian density
and investigate whether this (quantum-mechanical) Dirac equation is form
invariant under the actions of $\cP$ and $\cT$. This enables us to identify the
transformation properties of the interaction term and also to calculate the
dispersion relation associated with plane-wave solutions of the Dirac equation.
In addition, by rewriting the Dirac equation in the form $i\partial_t\psi=H
\psi$, we identify the {\it effective Hamiltonian} $H$ \cite{R18} associated
with the interaction. We will see that the form invariance of the Dirac equation
under $\cPT$ is equivalent to the statement that $H$ commutes with $\cPT$.

In analyzing the case of 1+1 dimensions, we find the surprising result that for
complex fermionic fields, the bilinear interaction form $-g\bar\psi\gamma^5\psi$
gives a Dirac equation that is {\it odd} under time reversal and also odd under
parity. Thus, the Dirac equation with the interaction term is form invariant
under $\cPT$. The $\cPT$ symmetry can also be verified by determining the
Hamiltonian $H$ associated with this Dirac equation $i\partial_t\psi=H\psi$. The
$2\times 2$ matrix representation clarifies this result. Comparing with the
general result for a $2\times2$ $\cPT$-symmetric fermionic Hamiltonian in 1+1
dimensions \cite{R19}, it becomes evident that in 1+1 dimensions the $\cPT$
symmetry is broken when $m$ vanishes. In a second example, due to the
similarities in the transformation properties of this interaction with those of
$\phi(t,{\bf x})$, we surmise that higher integer powers of the interaction
Lagrangian density $-g(\bar\psi\gamma^5\psi)^N$ might lead to a spectral
relation that has real energies; we investigate this for $N=2$ and 3. We
find that the $\cPT$ symmetry is always broken if we assume that the expectation
value $\langle\bar\psi\gamma^5 \psi\rangle$ is negative imaginary. There are no
other matrix potentials in 1+1 dimensions.

We then turn to further examples for which $x$-dependent $\cPT$-symmetric
potentials $ix^3$, $-x^4$, and $i\kappa/x$ introduced via vector or scalar
coupling various combinations, as well as the complex $\cPT$-symmetric lattice
potentials $i\kappa\cot(x)+i\gamma^0\sin(x)$ and the Hulth\'en potential are
included in the Dirac equation of motion. In order to gain some understanding of
these systems, we construct the analogous classical systems for which a
classical phase structure can be obtained. 

The situation in 3+1 dimensions is different because $\cT^2=-1$. Studying the
algebra in 3+1 dimensions, we confirm that the interaction term $-g\gamma^5\psi$
in the equation of motion is {\it even} under time reversal. Since the parity
transformation is still odd in 3+1 dimensions, we conclude that the interaction
term in the Dirac equation is not invariant under $\cPT$. While the dispersion
relation is superficially the same as for the 1+1-dimensional case, which
implies that there is {\it no} region in which the spectrum is real in the
chiral limit, the associated interaction Hamiltonian is anti-$\cPT$ symmetric,
which is consistent with the complex nature of the spectrum.

In 3+1 dimensions, we search for other bilinear combinations of fermionic fields
with the aim of determining all possible combinations that give a Dirac equation
that is form invariant under $\cPT$ and that are not Hermitian. We find two
types of terms having either an axial vector or a tensor structure. The spectra
of both of the non-Hermitian $\cPT$-symmetric interactions are analyzed. Here
too we find that the $\cPT$ symmetry is always broken in the chiral limit. We
also look at the consequences of imposing an additional condition that the
Hamiltonian be selfadjoint under the $\cPT$ inner product for fermions
\cite{r13,r12} and investigate the restrictions that this implies. We
demonstrate that the $\cPT$ symmetry is always broken in the chiral limit, a
feature that prevails in the analysis of the Dirac equation in the dimensions
studied. 

This paper is organized as follows: In Sec.~\ref{s2} we investigate possible
$\cPT$-symmetric interactions in 1+1 dimensions. We analyze $\cL_{\rm int}=-g
\bar\psi\gamma^5\psi$ in Subsec.~\ref{s2a} and extensions to this as $-g(\bar
\psi\gamma^5\psi)^N$ in Subsec.~\ref{s2b}. We introduce the spatially dependent
potentials $ix^3$, $-x^4$, and $i\kappa/x$, and the lattice and Hulth\'en
potentials in Subsec.~\ref{s2c}. In Sec.~\ref{s3} we analyze 3+1-dimensional
interactions, starting with $\cL_{\rm int}=-g\bar\psi\gamma^5\psi$ in
Subsec.~\ref{s3a} and other two-body (four-point) interactions in
Subsec.~\ref{s3b}. Our conclusions and outlook are presented in Sec.~\ref{s4}.

\section{Non-Hermitian $\cPT$-symmetric fermions in 1+1 dimensions}\label{s2}
\subsection{Axial bilinear fermionic interaction}\label{s2a}
We start with the Lagrangian density for a conventional Hermitian free fermionic
field theory,
\begin{equation}
\cL_0=\bar\psi(i\slashed{\partial}-m)\psi,
\label{E5}
\end{equation}
where $\bar\psi=\psi^\dagger\gamma^0$ and $\psi^\dagger$ is the Hermitian
conjugate of $\psi$. In 1+1 dimensions the gamma matrices are given in
(\ref{E1}). We have shown that the free Dirac equation (\ref{E2}) associated
with (\ref{E5}) is form invariant under the operation of $\cP$ in (\ref{E2a})
and of $\cT$ in (\ref{E2b}). Note that (\ref{E2}) is also form invariant under
the combined operations of $\cP$ and $\cT$ because the functions $\psi(t,x)$ and
$\cPT\psi(t,x)=\gamma^0\gamma^0\psi^*(-t,-x)=\psi^*(-t,-x)$ both satisfy
(\ref{E2}). A plane-wave solution to (\ref{E2}) gives the dispersion relation
$E^2=p^2+m^2$. Finally, we read off the {\it effective} or {\it
quantum-mechanical Hamiltonian} $H$ from the free Dirac equation $i\partial_t
\psi=H\psi$ in (\ref{E2}): $H=-i\gamma^0\gamma^1\partial_1+m\gamma^0$. (This
form is often written using the definitions $\alpha=\gamma^0\gamma^1$ and $\beta
=\gamma^0$ \cite{R18}).

We observe that the form invariance of the Dirac equation under $\cPT$ is
equivalent to the statement that $H$ commutes with $\cPT$: $H(\cPT\psi)=\cPT(H
\psi)$. This is so because the left hand side is
$$H(\cPT\psi)=H(\gamma^0\gamma^0\psi^*)=H\psi^*,$$
and the right hand side is
$$\cPT(H\psi)=\gamma^0\gamma^0\big(-i\gamma^0\gamma^1\partial_1+m\gamma^0\big)
\psi^*=H\psi^*.$$

Next, we examine what happens if a pseudoscalar bilinear term in included in the
Lagrangian density $\cL=\cL_0+\cL_{\rm int}$, where $\cL_{\rm int}=-g\bar\psi
\gamma^5\psi$ and $g$ is a real parameter. Now the associated quantum-mechanical
Dirac equation is altered to read
\begin{equation}
(i\slashed\partial-m-g\gamma^5)\psi=0.
\label{E8}
\end{equation}
Parity transforms this equation into
\begin{equation}
(i\slashed\partial-m+g\gamma^5)\gamma^0\psi(t,-x)=0,
\label{E9}
\end{equation}
and time reversal has the effect
\begin{equation}
(i\slashed\partial-m+g\gamma^5)\gamma^0\psi^*(-t,x)=0.
\label{E10}
\end{equation}
This Dirac equation is not invariant under $\cP$ or $\cT$ separately but it is
invariant under $\cPT$ because the axial interaction term changes sign twice; it
is odd under both $\cP$ and $\cT$. So this axial non-Hermitian term is $\cPT$
symmetric. 

We can formulate this differently: We identify the effective quantum-mechanical
Dirac Hamiltonian associated with the Dirac equation as
$$H=H_0+H_{\rm int}=-i\gamma^0\gamma^1\partial_1+m\gamma^0+g\gamma^0\gamma^5,$$
where $H_0=-i\gamma^0\gamma^1\partial_1+m\gamma^0$ and $H_{\rm int}=g\gamma^0
\gamma^5$. We have shown that $H_0$ commutes with $\cPT$, and from the effect of
$\cP$ and $\cT$ in 1+1 dimensions and the reality of $H_{\rm int}$, we see that
$H_{\rm int}$ also commutes with $\cPT$. Thus, the effective Hamiltonian $H$
reflects the symmetry of the Dirac equation.

For this case the dispersion relation is obtained from a plane-wave solution
$\psi(t,x)$, and multiplying (\ref{E8}) by $(\slashed p+m+g\gamma^5)$, where
$\slashed p=\gamma^0p_0+\gamma^1 p_1$, yields the result \cite{r15} $p^2=m^2-
g^2$, which is positive only when $m^2\geq g^2$. Thus, in the chiral limit $m\to
0$ the spectrum is complex and the $\cPT$ symmetry is broken in this limit.

The matrix representation makes this result clearer. Recall that a general
two-dimensional $\cPT$-symmetric fermionic Hamiltonian, which is selfadjoint
with respect to the $\cPT$ inner product for fermions and which commutes with
$\cPT$, can be written as \cite{R19}
\begin{equation}
H^{\cPT}=\begin{pmatrix}a & b\\ f & a \end{pmatrix},
\label{E11}
\end{equation}
where $a$, $b$, and $f$ are real numbers. The eigenvalues are $E_{\pm}=a\pm\sqrt
{bf}$. Thus, if $b$ and $f$ have the same sign, the spectrum is real and the
$\cPT$ symmetry is unbroken.

Now, if the interaction Lagrangian density is $-g\bar\psi\gamma^5\psi$, the
quantum-mechanical interaction Hamiltonian is
$$H_{\rm int}=g\gamma^0\gamma^5=g\begin{pmatrix} 0 & 1\\ -1 & 0\end{pmatrix}.$$
Comparing this with (\ref{E11}), we confirm that $H_{\rm int}$ is $\cPT$
symmetric and that this symmetry is always broken. Note that $H_{\rm int}$ is
non-Hermitian.

If we add the conventional mass term to the interaction, the effective
Hamiltonian in matrix form becomes
$$H=m\gamma^0+g\gamma^0\gamma^5=\begin{pmatrix}0 & m+g\\ m-g & 0\end{pmatrix}.$$
We see immediately that it is $\cPT$ symmetric and that the $\cPT$ symmetry is
unbroken if $g^2\leq m^2$.

Observe that the equation of motion resulting from the Dirac equation with an
{\it imaginary} axial term, 
\begin{equation}
(i\slashed\partial-m-ig\gamma^5)\psi=0,
\label{E12}
\end{equation}
gives the dispersion relation $p^2=m^2+g^2$. So $m$ is real for all $g$,
including the chiral limit $m\to0$. However, this axial term is not $\cPT$
symmetric. In fact, it is {\it anti}-$\cPT$ symmetric, so that (\ref{E12}) is
not form invariant under $\cPT$.

\subsection{Approximate solution for higher-power field-theoretic interactions}
\label{s2b}
This section explores the effect that higher-power interaction terms have in
$1+1$-dimensional systems. Our starting point is the general Lagrangian density
$$\cL(N)=\cL_0+\cL_{\rm int}=\bar\psi(i\slashed{\partial}-m)\psi-g(\bar\psi
\gamma^5\psi)^N.$$
The Euler-Lagrange equations give the corresponding Dirac or single-particle
equation of motion as 
$$\big[i\slashed\partial-m-Ng\gamma^5(\bar\psi\gamma^5\psi)^{N-1}\big]\psi=0,$$
which is nonlinear if $N>1$. The case $N=1$ reduces to that examined in
Sec.~\ref{s2a}. In the following we examine the cases $N=2$ and $N=3$.

\subsubsection{$N=2$}
When $N=2$, $\cL(2)=\bar\psi(i\slashed{\partial}-m)\psi-g(\bar\psi\gamma^5\psi
)^2$, and the interaction term is $\cPT$ symmetric. The associated
Euler-Lagrange equation is
\begin{equation}
\big[i\slashed\partial-m-2g\gamma^5(\bar\psi\gamma^5\psi)\big]\psi=0,
\label{E13}
\end{equation}
from which we deduce that the Hamiltonian $H$ satisfying $i\partial_t\psi=H\psi$
is
$$H=-i\gamma^0\gamma^1\partial_1+m\gamma^0+2g\gamma^0\gamma^5(\bar\psi\gamma^5
\psi).$$
To solve (\ref{E13}) approximately we replace $\bar\psi\gamma^5\psi$ by its
average value $\langle\bar\psi\gamma^5\psi\rangle=\langle\phi\rangle$.
Furthermore, by operating on the approximate version of (\ref{E13}) by
$(i\slashed\partial-m-2g\gamma^5\langle\phi\rangle)$ we can solve for the
spectrum. In the chiral limit $m\to0$ this is
\begin{equation}
p^2=-4g^2\langle\phi\rangle^2.
\label{E14}
\end{equation}
Now, noting that the expectation value of a bosonic pseudoscalar field should be
negative imaginary \cite{r10}, 
$$\langle\phi\rangle=-iA,$$
where $A$ is a constant, it follows from (\ref{E14}) that $p^2=4g^2A^2$ is real.
However, with this choice of $\langle\phi\rangle$, $H_{\rm int}= 2g\gamma^0
\gamma^5\langle\phi\rangle$ is anti $\cPT$ symmetric, as is the interaction term
in (\ref{E13}). Thus, the quantum-mechanical Dirac equation is no longer form
invariant under $\cPT$; also $\cPT$ does not commute with $H$. Yet we obtain a
real spectrum because now $H_{\rm int}$ is Hermitian. The opposite case, namely,
when the Dirac equation is $\cPT$ symmetric and $H$ commutes with $\cPT$, can be
simulated by letting $g\to ig$. Then $p^2<0$ so, as in Sec.~IIA, $\cPT$ symmetry
is again realized in the broken phase.

\subsubsection{$N=3$}
When $N=3$, $\cL(3)=\bar\psi(i\slashed{\partial}-m)\psi-g(\bar\psi\gamma^5\psi
)^3$. This resembles the case for $N=1$. The Euler-Lagrange equation now reads
\begin{equation}
\big[i\slashed\partial-m-3g\gamma^5(\bar\psi\gamma^5\psi)^2\big]\psi=0.
\label{E15}
\end{equation}
It follows that the interaction part of the Hamiltonian is
\begin{equation}
H_{\rm int}=3g\gamma^0\gamma^5(\bar\psi\gamma^5\psi)^2.
\label{E16}
\end{equation}
Again, to find an approximate solution we replace $(\bar\psi\gamma^5\psi)^2$ by
its average value $\langle(\bar\psi\gamma^5\psi)^2\rangle$. Solving (\ref{E15})
we get
$$p^2=-9g^2\langle(\bar\psi\gamma^5\psi)^2\rangle^2$$
in the chiral limit. We expect $\langle(\bar\psi\gamma^5\psi)^2\rangle$ to be
real, so $p^2<0$ and the $\cPT$ symmetry is always broken. We can confirm this
explicitly by noting that (\ref{E16}) is simply proportional to $\gamma^1$ and
thus only has off-diagonal values of opposite sign, see (\ref{E1}). Comparing
this with (\ref{E11}), we note that (\ref{E16}) is manifestly $\cPT$ symmetric.

We conclude that (i) If we construct a 1+1-dimensional Lagrangian density
containing the axial $\cPT$-symmetric interaction $(\bar\psi\gamma^5\psi)^N$
($N$ odd), our approximation scheme shows that we obtain an equation of motion
that is form invariant under $\cPT$, and correspondingly a $\cPT$-symmetric
Hamiltonian. The $\cPT$ symmetry is broken in the chiral limit. (ii) For even
$N$ the equation of motion contains an anti-${\cPT}$-symmetric term and the
associated interaction Hamiltonian is also anti-$\cPT$ symmetric but we obtain a
dispersion relation that has real masses as a result of Hermiticity. If we
modify the interaction by replacing $g\to ig$, we obtain a $\cPT$-symmetric
system but once again the $\cPT$ symmetry is broken.

\subsection{Dirac particle in $\cPT$-symmetric potentials}\label{s2c}
In 1+1 dimensions there are no other $\gamma$-matrix-based interactions.
However, in addition to these, we can include $\cPT$-symmetric potentials having
a spatial dependence such as $ix^3$, $-x^4$, $i\kappa/x$, or even periodic
potentials into the relativistic Dirac equation and study the effects of these.
Unlike nonrelativistic potentials, which are scalars and can only be included as
such in the Schr\"odinger equation, in the Dirac equation, such potentials can
be incorporated either as the nonvanishing scalar part of the 4-vector
potential (which we refer to as vector coupling), or as pure scalar
interactions, or as combinations thereof. We consider some examples below.

\subsubsection{Vector coupling with $ix^3$}
\begin{figure}
\centering
\includegraphics[scale=0.65]{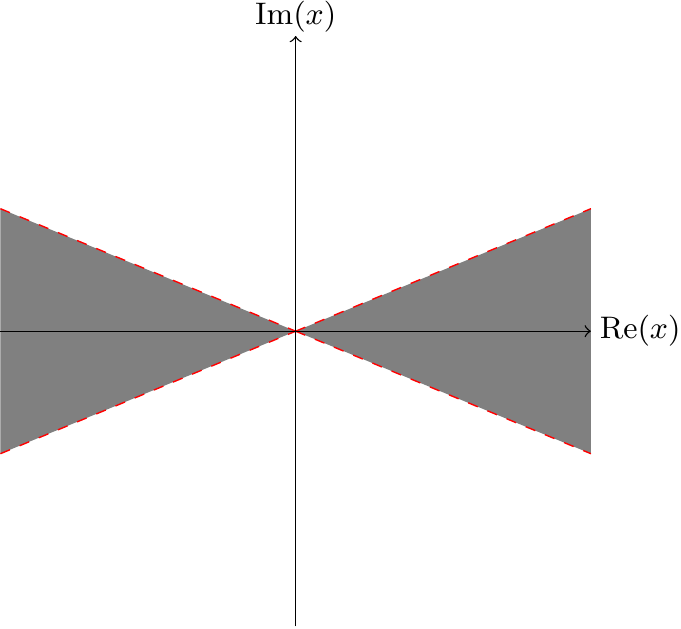}
\caption{Stokes sectors in the complex-$x$ plane for $\psi_1$ with an opening
angle of $\pi/4$ for the massless Dirac particle in the vector coupled potential
$ix^3$; $\psi_1$ vanishes exponentially as $|x|\to\infty$ inside these sectors.}
\label{f1}
\end{figure}

\begin{figure}
\centering
\includegraphics[scale=0.65]{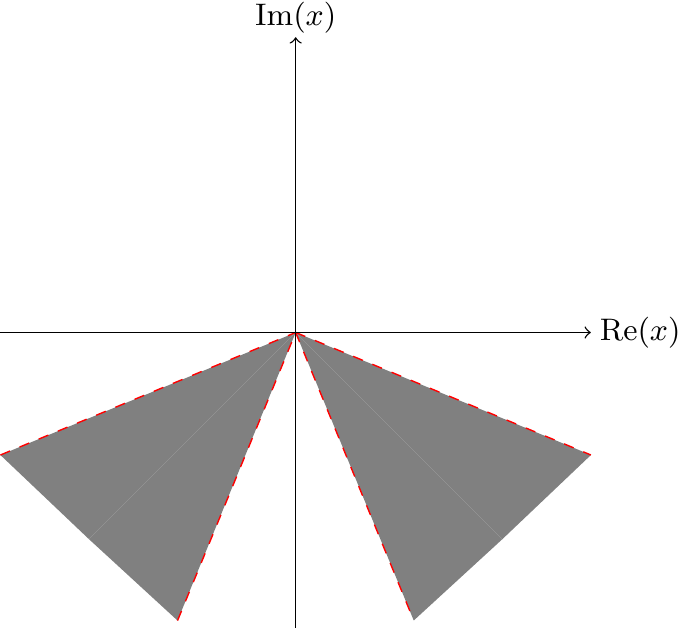}
\caption{Stokes sectors in the complex-$x$ plane for $\psi_2$ with an opening
angle of $\pi/4$. In this case, the sectors rotate below the real-$x$ axis;
$\psi_2$ vanishes exponentially as $|x|\to\infty$ inside these sectors.}
\label{f2}
\end{figure}

The 1+1-dimensional Dirac equation that includes the non-Hermitian
$\cPT$-symmetric vector-coupled potential $ix^3$ reads
\begin{equation}
\big(i\slashed\partial-ix^3\gamma^0\big)\psi(t,x)=0.
\label{edix}
\end{equation}
This is form invariant under $\cPT$ and the associated relativistic Hamiltonian 
$$H=-i\alpha \partial_x+ix^3\quad(\alpha\equiv\gamma^0\gamma^1),$$
is also $\cPT$ invariant. If we look for solutions of the form $\psi(t,x)=e^{-i
Et}\psi(x)$, we arrive at the corresponding eigenvalue problem 
$$H\psi=\big(-i\alpha\partial_x+ix^3\big)\psi=E\psi.$$
The eigenvectors $\psi_1(x)$ and $\psi_2(x)$ that solve this equation have the
asymptotic behavior
$$\psi_1(x)\sim\left(\begin{array}{cc}e^{-x^4/4}\\ 0\\ \end{array}\right),
\qquad\psi_2(x)\sim\left(\begin{array}{cc} 0\\ e^{x^4/4}\end{array}
\right).$$
The convergence domain for $\psi_1(x)$ and $\psi_2(x)$ in the complex-$x$ plane
are the $\cPT$-symmetric Stokes sectors shown in Figs.~\ref{f1} and \ref{f2}
respectively. In these sectors $\psi_{1,2}(x)$ vanish exponentially as $|x|\to
\infty$.

To obtain the selfenergy of the propagating particle we apply $i\slashed
\partial$ to (\ref{edix}) and obtain the differential equation
\begin{equation}
\big(E^2+\partial_x^2\big)\psi=-\big(x^6+3x^2\gamma^1\gamma^0+2x^3\partial_x
\gamma^1\gamma^0\big)\psi.
\label{e13}
\end{equation}
Since the matrix $\gamma^1\gamma^0={\rm diag}(1,-1)$ is diagonal, the
two-component equations in (\ref{e13}) decouple. Although they are not
Schr\"odinger-like, each is individually $\cPT$ symmetric. We first examine the
classical analog of these equations obtained by replacing $-i\partial_x$ by $p$,
\begin{eqnarray}
&&\left(\begin{array}{cc} E^2-p^2 & 0\\ 0 & E^2-p^2\\ \end{array}\right)=
\nonumber\\
&& \quad\left(\begin{array}{cc} -x^6-2ipx^3-3x^2 & 0\\ 0 & -x^6+2ipx^3+3x^2\\
\end{array}\right).\nonumber
\end{eqnarray}

\begin{figure}
\centering
\includegraphics[scale=0.5]{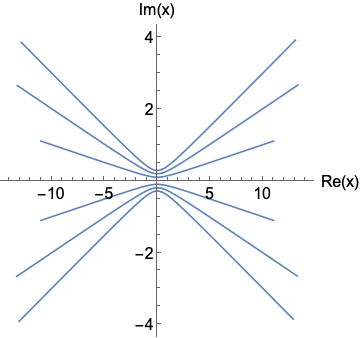}
\caption{Classical trajectories in the complex-$x$ plane described by $H_1=\sqrt
{p^2-2ix^3p-x^6-3x^2}$.}
\label{f3}
\end{figure}

The classical Hamiltonian associated with $\psi_1$ is $H_1=\sqrt{p^2-2ix^3p-x^6
-3x^2}$. The equation of motion of a classical particle described by $H_1$ is
obtained by combining Hamilton's equations $dx/dt=\partial H_1/\partial p$ and
$dp/dt=-\partial H_1/\partial x$: $dx/dt=\pm\sqrt{1+3x^2/E^2}$. By rescaling
both $x$ and $t$ this equation becomes
$$\frac{dx}{dt}=\pm\sqrt{1+x^2}.$$
We find that $x(t)$ forms {\it open} trajectories in the complex-$x$ plane, as
shown in Fig.~\ref{f3}.

The open classical trajectories of the particle in the complex-$x$ plane
reflects the behavior seen in the quantum case: By setting $p=0$, we observe
that the selfenergy $\Sigma_1$ of the particle corresponding to $\psi_1$ is
given by $\Sigma^2_1=-x^6-3x^2$, which implies that $\Sigma_1$ cannot be real 
\cite{BBH}.

\begin{figure}
\centering
\includegraphics[scale = 0.5]{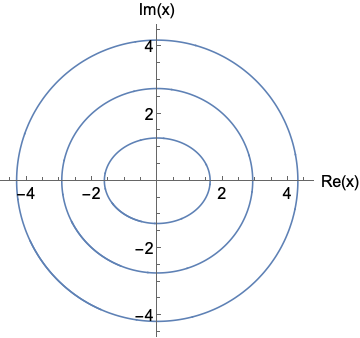}
\caption{Classical trajectories in the complex-$x$ plane described by $H_2=\sqrt
{p^2+2ix^3p-x^6+3x^2}$.}
\label{f4}
\end{figure}

On the other hand, the trajectories of the classical particle in the complex-$x$
plane that are associated with the classical Hamiltonian $H_2=\sqrt{p^2+2ix^3p-
x^6+3x^2}$, are closed, as can be seen in Fig.~\ref{f4}. In the quantum system,
the selfenergy corresponding to $\psi_2$ is $\Sigma^2_2=-x^6+3x^2$. By
parametrizing $x$ as $-i(\sqrt{1+ir}-1)$ where $r$ is real, $\psi_2$ vanishes
exponentially as $r\to\pm\infty$. We note that the ends of this path lie in the
left and right Stokes sectors of Fig.~\ref{f2} as $|x|\to\infty$. When $-\sqrt
[4]{3}<x<\sqrt[4]{3}$, $\Sigma^2_2$ is positive. Thus, the selfenergy associated
with the particle is real.

\subsubsection{Scalar coupling with $ix^3$ and vector coupling with $i\kappa/x$}
In the previous subsection we treated the $\cPT$-symmetric potential $ix^3$ in a
vector-coupling scheme; now we consider it as a {\it scalar} potential, where,
in addition, the Dirac particle is also under the influence of a complex
$\cPT$-symmetric Coulomb potential. The non-Hermitian $\cPT$-symmetric Dirac
equation now reads
\begin{equation}
\big(i\slashed\partial-(i\kappa/x)\gamma^0-ix^3\big)\psi(t,x)=0,
\label{ed2}
\end{equation}
where $\kappa$ is a real parameter. The associated relativistic quantum-mechanical Hamiltonian is
$$H=-i\alpha\partial_x+i\kappa/x+\beta ix^3\quad\big(\beta\equiv\gamma^0\big).$$
Again, looking for solutions of the form $\psi(t,x)=e^{-iEt}\psi(x)$ leads to an
eigenvalue problem
$$H\psi=\big(-i\alpha\partial_x+i\kappa/x+\beta ix^3\big)\psi=E\psi.$$
Writing the eigenfunction $\psi(x)$ in terms of its two spinor components, $\psi
(x)=(\phi_1(x),\phi_2(x))$ \footnote{We reserve the notation $\psi_1,\psi_2$ for
two different spinor solutions of the same equation.} we find two coupled
differential equations for the scalar functions $\phi_{1,2}(x)$,
\begin{eqnarray}
i\phi_1'+i\kappa\phi_1/x+ix^3\phi_2&=&E\phi_1, \label{e14}\\
-i\phi_2'+i\kappa\phi_2/x+ix^3\phi_1&=&E\phi_2. \label{e15}
\end{eqnarray}

We can eliminate the second component $\phi_2$ from (\ref{e14}) by exploiting
(\ref{e15}), and after rescaling $\phi_1$, and choosing $\kappa$ to be $-3/2$
for convenience, we obtain the simple form
\begin{equation}
-\phi_1''-x^6\phi_1=E^2\phi_1,
\label{e16}
\end{equation}
which is a Schr\"odinger-like equation with a $-x^6$ potential. On the real-$x$
axis this upside-down potential is unstable, but by imposing appropriate
$\cPT$-symmetric boundary conditions we can obtain a real spectrum. As in the
previous subsection, we find that to have a convergent eigenfunction, we must
treat the problem in the complex-$x$ plane.

\begin{figure}
\centering
\includegraphics[scale=0.9]{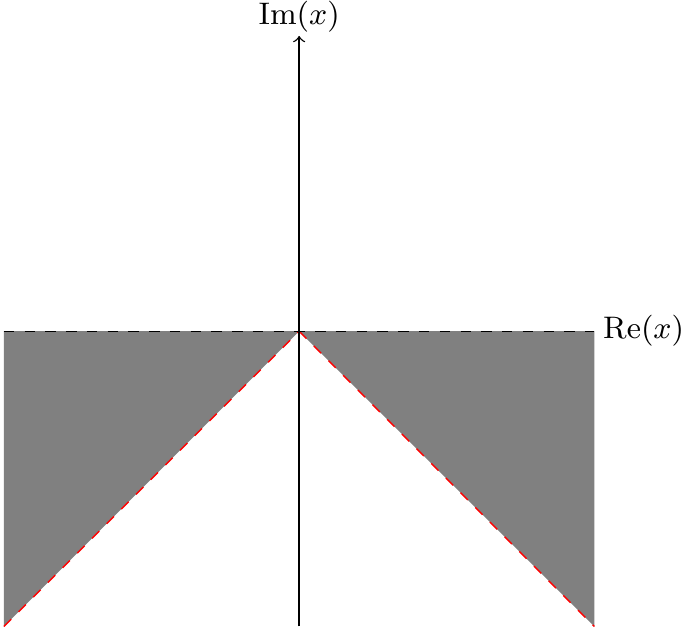}\setlength{\unitlength}{0.14in}
\caption{Stokes sectors in the complex-$x$ plane for $\phi_1$ in (\ref{e16}).
$\phi_1$ vanishes exponentially inside these sectors.}
\label{f5}
\end{figure}

The WKB approximation for the solutions of (\ref{e16}) to leading order is
\cite{cmbbook}
\begin{equation}
\phi_{\rm WKB}(x)=C_\pm[Q(x)]^{-1/4}e^{\pm i\int^x\,ds\sqrt{Q(s)}},
\end{equation}
where $Q(x)=E^2+x^6$. For large $|x|$ the exponential component of this
asymptotic behavior is
\begin{equation}
\phi_1\sim e^{\pm ix^4/4}.
\label{e17}
\end{equation}
There are eight Stokes sectors in the complex-$x$ plane, each with an opening
angle of $\pi/4$. To have a $\cPT$-symmetric pair of Stokes sectors, we choose
the minus sign in (\ref{e17}) for the right Stokes sector, which is located just
below the positive-real-$x$ axis. For the left Stokes sector we choose the
positive sign in (\ref{e17}), which determines a sector located just below the
negative-real-$x$ axis. These two Stokes sectors are depicted in Fig.~\ref{f5}.

We can also approximate the eigenenergies of (\ref{e16}). To do so, we first
find the two turning points which are determined by $E=- x^6$ and which lie in
the Stokes sectors in Fig.~\ref{f5}. These two points are
$$x_1=\sqrt[6]{E}e^{- 5i\pi/6},\qquad x_2=\sqrt[6]{E}e^{-i\pi/6}.$$
The WKB quantization condition is
$$\int_{x_1}^{x_2}ds\sqrt{E_n^2+s^6}=\big(n+\half\big)\pi\quad(n\to\infty).$$
Thus,
$$E_n=\pm\Big[4\sqrt{\pi/3}\Gamma(\textstyle{\frac{2}{3}})(2n+1)/\Gamma(
\textstyle{\frac{1}{6}})\Big]^{3/4}\quad(n\to\infty).$$
For $n=0$ or $1$, we obtain $E_0=\pm 1.0$ and $E_1=\pm 2.27$. 

\begin{figure}
\centering
\includegraphics[scale=0.5]{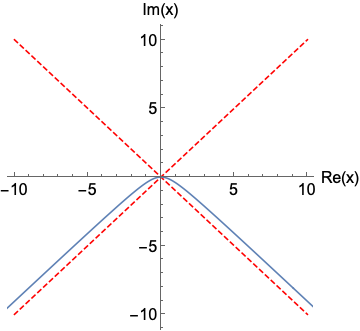}
\caption{The contour (solid line) on which the eigenvalue problem in (\ref{e16})
is posed (blue online). The dashed lines (red online) denote the edges of the
sectors.}
\label{f6}
\end{figure}

An exact calculation of the eigenvalues can be made on parametrizing $x$ as $-i
(\sqrt{1+ir}-1)$, where $r$ is a real variable. As depicted in Fig.~\ref{f6},
the ends of this path lie inside the Stokes sectors as $|x|\to\infty$, so we
pose the eigenvalue problem for the differential equation in (\ref{e16}) on this
contour. We determine the ground-state and first-excited-state energies
numerically as
$$E_0=\pm1.16,\qquad E_1=\pm2.29,$$
which illustrates the accuracy of the WKB approximation. Thus, the energy
spectrum of the Dirac particle in the combined non-Hermitian $\cPT$-symmetric
potentials $ix^3$ and $i\kappa/x$ is real and discrete.

\begin{figure}
\centering
\includegraphics[scale=0.5]{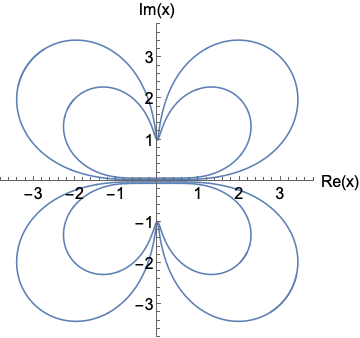}
\caption{Classical trajectories in the complex-$x$ plane described by $H=\sqrt{
p^2-x^6}$.}
\label{f7}
\end{figure}

The trajectories of a classical particle in the complex-$x$ plane described by
the classical Hamiltonian $H=\sqrt{p^2-x^6}$ obtained from (\ref{e16}) are shown
in Fig.~\ref{f7}. These trajectories are closed, which reflects the reality and
the discreteness of the spectrum at the quantum level.

\subsubsection{Vector coupling with $-x^4$} 
Next, we consider a massless Dirac particle under the influence of the
upside-down quartic potential $-x^4$. In the vector-coupling scheme, the
relativistic Dirac equation is modified to read
\begin{equation}
(i\slashed \partial+x^4\gamma^0)\psi(t,x)=0.
\label{ed3}
\end{equation}
As in the previous examples, this equation is form invariant under $\cPT$ and
the associated Hamiltonian
$$H=-i \alpha \partial_x-x^4,$$
commutes with $\cPT$. Looking for solutions of the form $\psi(t,x)=e^{-iEt}\psi
(x)$ leads to an eigenvalue equation $H\psi=E\psi$, whose eigenvectors behave
asymptotically as
$$\psi_1\sim\left(\begin{array}{cc}e^{-ix^5/5}\\ 0\\ \end{array}\right),
\qquad\psi_2\sim\left(\begin{array}{cc} 0\\ e^{ix^5/5}\end{array}\right).$$
Note that $\psi_1$ vanishes exponentially in a Stokes sector with opening angle
$\pi/5$. This sector contains the negative-imaginary-$x$ axis, so it vanishes
exponentially as $x\to-i\infty$. The function $\psi_2$ also vanishes
exponentially in the same Stokes sector, but one that has rotated upward; that
is, $\psi_2\to0$ as $x\to i\infty$.

Following the analysis given in Subsec.~1, we iteratively apply $i\slashed
\partial$ to the corresponding Dirac equation and find the decoupled system of
equations
\begin{eqnarray}
&&\left(\begin{array}{cc} E^2-p^2 & 0\\ 0 & E^2-p^2\\ \end{array}\right)=
\nonumber\\
&& \quad\left(\begin{array}{cc} x^8+2px^4-4ix^3 & 0\\ 0 & x^8-2px^4+4ix^3\\
\end{array}\right).
\label{e18}
\end{eqnarray}
The selfenergies $\Sigma_1$ and $\Sigma_2$ of the particle corresponding to
$\psi_1$ and $\psi_2$ are given by $\Sigma^2_1=x^8-4ix^3$ and $\Sigma^2_2=x^8+4i
x^3$. As $\psi_1$ and $\psi_2$ converge on $x=-ir$ and $x=ir$, the selfenergies
become real.

\begin{figure}
\centering
\includegraphics[scale=0.5]{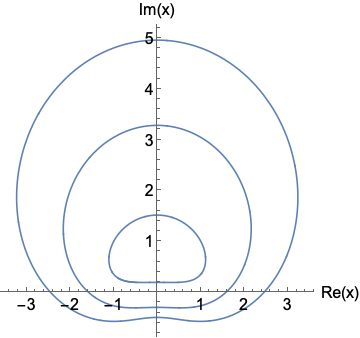}
\caption{Classical trajectories in the complex-$x$ plane described by $H=\sqrt{
p^2+2x^4p+x^8-4ix^3}$.}
\label{f8}
\end{figure}

The trajectories of the classical particle described by both of the classical
Hamiltonians obtained from (\ref{e18}) are closed in the complex-$x$ plane. In
Fig.~\ref{f8}, this is shown for the classical Hamiltonian $H=\sqrt{p^2+2x^4p+
x^8-4ix^3}$.

\subsubsection{Scalar coupling with $-x^4$ and vector coupling with $i\kappa/x$}
We now treat the upside-down potential $-x^4$ as a scalar potential and, in
addition, we consider the effect of a complex $\cPT$-symmetric Coulomb potential
on the Dirac particle in a vector-coupling scheme, satisfying the modified Dirac
equation
$$\big(i\slashed\partial-(i\kappa/x)\gamma^0+x^4\big)\psi(t,x)=0,$$
where $\kappa$ is a real parameter. This equation is form invariant under $\cPT$
and the associated Hamiltonian
$$H=-i\alpha \partial_x+i\kappa/x-\beta x^4,$$
commutes with $\cPT$. The search for solutions of the form $\psi(t,x)=e^{-iEt}
\psi(x)$ requires solutions of the eigenvalue equation
$$H\psi=\big(-i\alpha \partial_x+i\kappa /x-\beta x^4\big)\psi=E\psi.$$
As in Subsec.~2, it is convenient to write $\psi(x)$ in terms of its (scalar)
components, $\psi=(\phi_1,\phi_2)$ and derive the coupled equations that
$\phi_1$ and $\phi_2$ satisfy. Following the procedure outlined in Subsec.~2, we
eliminate $\phi_2$ and arrive at a Schr\"odinger-like equation for $\phi_1$,
\begin{equation}
-\phi_1''+x^8\phi_1=E^2\phi_1,
\label{e19}
\end{equation}
where, for convenience, we have set $\kappa=-2$. We have thus found an octic
potential with positive sign. Hence, we pose the eigenvalue problem on the
real-$x$ axis. As before we use the WKB approximation to obtain the
eigenvalues for large $n$,
$$E_n=\pm\Big[\sqrt\pi\Gamma(\textstyle{\frac{13}{8}})\big(n+\half\big)/\Gamma(
\textstyle{\frac{9}{8}})\Big]^{4/5}\quad(n\to\infty).$$
From this equation we find that $E_0=\pm0.87$ and $E_1=\pm2.10$. A direct
numerical calculation gives $E_0=\pm1.11$ and $E_1=\pm2.18$. Thus, once again,
we find that the energy spectrum of a Dirac particle in the presence of combined
non-Hermitian $\cPT$-symmetric vector and scalar potentials $i\kappa/x$ and
$-x^4$ is real and discrete.

\begin{figure}
\centering
\includegraphics[scale=0.5]{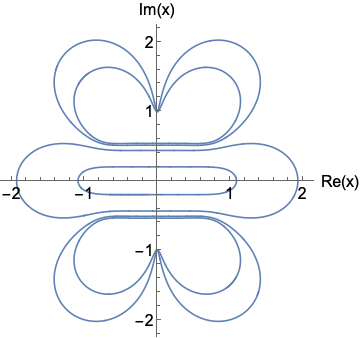}
\caption{Classical trajectories in the complex-$x$ plane described by $H=\sqrt{
p^2+x^8}$.}
\label{f9}
\end{figure}

Here again, we see that the reality and discreteness of the spectrum is evident
at the classical level with closed trajectories in the complex-$x$ plane. We
recognize the classical Hamiltonian of the system from (\ref{e19}) as being $H=
\sqrt{p^2+x^8}$. Figure~\ref{f9} shows that the classical trajectories
described by this Hamiltonian $H$ are closed.

\subsubsection{Complex $\cPT$-symmetric lattice potentials}
The methods in the previous subsections are general enough to be applied to a
Dirac particle in complex $\cPT$-symmetric lattices. The relativistic Dirac
equation
\begin{equation}
(i\slashed\partial-i\kappa\cot(x)\gamma^0-i\sin(x))\psi(t,x)=0,
\label{ed4}
\end{equation}
with $\kappa$ real, has non-Hermitian interaction terms, but is form invariant
with respect to $\cPT$. The associated Hamiltonian,
$$H=-i\alpha \partial_x+i\kappa\cot(x)+i\beta\sin(x),$$
commutes with $\cPT$.

As before, we can search for time-independent solutions of (\ref{ed4}). Writing
$\psi(t,x)=e^{-iEt}\psi(x)$, we obtain coupled equations for the components of
the spinor eigenfunction $\phi_1$ and $\phi_2$, where $\psi=(\phi_1,\phi_2)$.
Eliminating $\phi_2$, we find a Schr\"odinger-like equation for the $\phi_1$,
which after suitably rescaling, is 
$$-\phi_1''-\sin^2(x)\phi_1=E^2\phi_1,$$
where we have set $\kappa=-1/2$.

\begin{figure}
\centering
\includegraphics[scale=0.5]{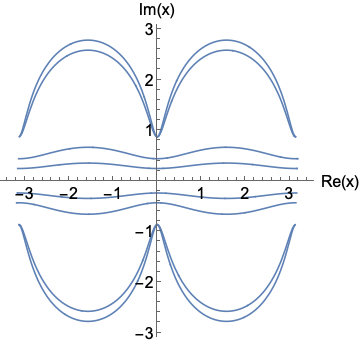}
\caption{Classical paths for $H=\sqrt{p^2-\sin^2(x)}$.}
\label{f10}
\end{figure}

The spectrum of the operator $-d^2/dx^2-\sin^2(x)$ is real and consists of
spectral bands separated by infinitely many spectral gaps \cite{r17}. The
absence of discrete energies and the reality of the band-structure manifest
itself via periodic, open trajectories of the classical particle described by
$H=\sqrt{p^2-\sin^2(x)}$ in the complex-$x$ plane, as depicted in
Fig.~\ref{f10}.

\begin{figure*}
\centering
\includegraphics[scale = 0.5]{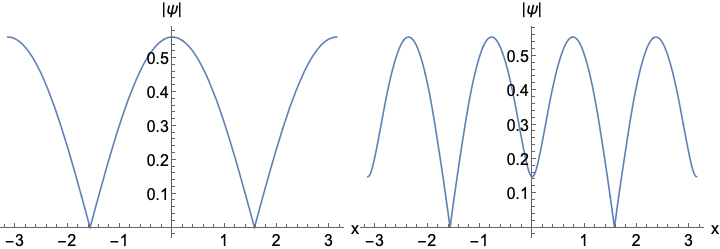}
\caption{Absolute values of the eigenfunctions corresponding to the band-edge
energies of $1.08$ (left panel) and $3.97$ (right panel) of $H=p^2+i\sin(x)$.}
\label{f11}
\end{figure*}

\begin{figure}
\centering
\includegraphics[scale = 0.6]{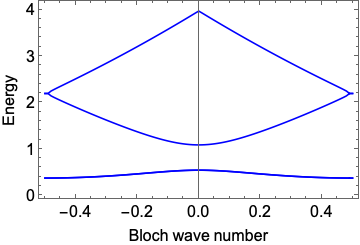}
\caption{The energy bands associated with the potential $i\sin(x)$ in the
first Brillouin zone.}
\label{f12}
\end{figure}

Before closing this subsection, we make a side remark: We note that the
(quantum-mechanical, nonrelativistic) Hamiltonian $H=p^2+i\sin(x)$ describes a
particle subject to the periodic potential $V(x)=i\sin(x)$ in a $\cPT$-symmetric
crystal. As was shown in Ref.~\cite{r18}, by examining a discriminant, one can
conclude that this Hamiltonian has real energy bands. However, to verify that
the band structure is real, one can alternatively show that the {\it
eigenfunctions} are $\cPT$ symmetric; that is, that the $\cPT$ symmetry of the
Hamiltonian is unbroken. To this end we plot the absolute values of the
eigenfunctions of the two states of $H=p^2+i\sin(x)$ in Fig.~\ref{f11} and
observe that both are in fact symmetric. The energy bands are real, and are
shown in Fig.~\ref{f12}. We use this technique in the next subsection.

\subsubsection{Scalar coupling with complex $\cPT$-symmetric Hulth\'en
potential}

\begin{figure*}
\centering
\includegraphics[scale=0.5]{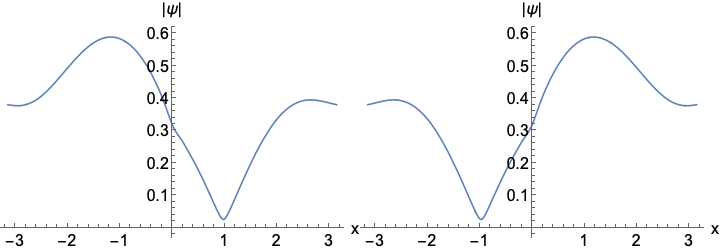}
\caption{Absolute values of the eigenfunctions corresponding to the band-edge
energies $0.75+i0.59$ (left panel) and $0.75-0.59i$ (right panel) for $H=p^2+
e^{-ix}/(1-e^{-ix})$.}
\label{f13}
\end{figure*}

The complex $\cPT$-symmetric Hulth{\'e}n potential is
$$V(x)=\frac{e^{-ix}}{1-e^{-ix}}.$$
If we regard $V(x)$ as a potential in the {\it nonrelativistic} time-independent
Schr\"odinger equation, $H\psi=E\psi$, with $H=p^2+V(x)$, we find that the band
structure for the energies is entirely complex, and, as is the case with
$\cPT$-symmetric potentials in the broken-symmetry phase, the eigenvalues occur
in complex-conjugate pairs. We illustrate this by plotting the absolute values
of the eigenfunctions of the two states of the Hamiltonian that correspond to
the complex-conjugate pairs of the band-edge energies $E=0.75\pm 0.59i$, see
Fig.~\ref{f13}. Note that the eigenfunctions display no symmetry, which implies
the complex nature of the band structure.

We now consider the relativistic Dirac equation that includes the
$\cPT$-symmetric Hulth{\'e}n potential in a scalar-coupling scheme, together
with an additional $\cPT$-symmetric vector potential:
\begin{equation}
\Big(i\slashed\partial-\kappa\frac{1}{1-e^{-ix}}\gamma^0-\frac{e^{-ix}}{1-e^{-ix
}}\Big)\psi(t,x)=0,
\label{ed5}
\end{equation}
with $\kappa$ being a real parameter. This equation has been constructed so as
to be form invariant with respect to $\cPT$ and the associated Hamiltonian
$$H=-i \alpha \partial_x+\kappa\frac{1}{1-e^{-ix}}+\beta\frac{e^{-ix}}{1-e^{-ix}
},$$
once again commutes with $\cPT$. Following the same procedure as in the last
subsections, we search for time-independent solutions of the Dirac equation, and
find the equations for the components of $\psi=(\phi_1,\phi_2)$. On eliminating
$\phi_2$, we obtain a Schr\"odinger-like equation for the first component of the
two-component spinor eigenfunction as
\begin{equation}
-\phi_1''+\frac{1}{(1-e^{ix})^2}\phi_1=E^2\phi_1,
\label{e20}
\end{equation}
where we have set $\kappa=-1/2$ for convenience.

\begin{figure*}
\centering
\includegraphics[scale = 0.5]{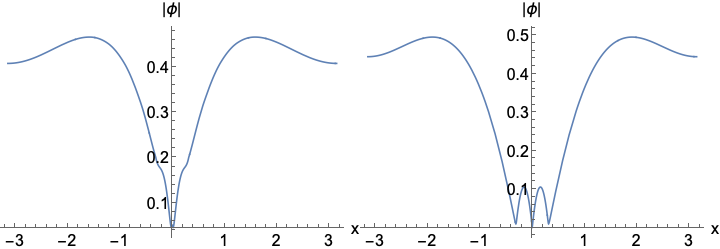}
\caption{Absolute values of the eigenfunctions corresponding to the band-edge
energies of $E=0.65$ (left panel) and $E=0.98$ (right panel), obtained from
(\ref{e20}). The symmetry of the eigenfunctions implies the reality of the
energy band.}
\label{f14}
\end{figure*}

\begin{figure}
\centering
\includegraphics[scale=0.6]{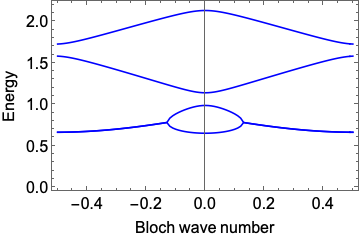}
\caption{The energy bands for the potential $(1-e^{ix})^{-2}$ in the first
Brillouin zone.}
\label{f15}
\end{figure}

By using spectral methods, we determine numerically that the band structure in
(\ref{e20}) is entirely real; that is, the symmetry is unbroken. We have shown
the absolute values of the first two eigenfunctions in Fig.~\ref{f14}, which
are clearly symmetric as expected. The (real) energy bands corresponding to this
potential are shown in Fig.~\ref{f15}.

\begin{figure}
\centering
\includegraphics[scale=0.5]{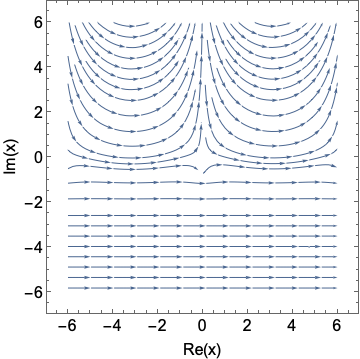}
\caption{Classical trajectories in the complex-$x$ plane described by $H=\sqrt{
p^2+1/(1-e^{i x})^2}$.}
\label{f16}
\end{figure}

The classical Hamiltonian associated with this system is $H=\sqrt{p^2+1/(1-
e^{ix})^2}$. The trajectories of the classical particle, as shown in
Fig.~\ref{f16}, are periodic and open. This appears to correspond to the fact
that the quantum Hamiltonian has real energy bands but no discrete eigenvalues. 

\section{Non-Hermitian $\cPT$-symmetric fermions in 3+1 dimensions}\label{s3}
\subsection{Axial bilinear fermionic interaction}
\label{s3a}
In 3+1 dimensions, we again start with the free fermionic Lagrangian $\cL_0=\bar
\psi(i\slashed{\partial}-m)\psi$ of (\ref{E5}) and the Dirac equation of motion,
\begin{equation}
(i\slashed\partial-m)\psi(t,{\bf x})=0,
\label{ee21}
\end{equation}
and recall that the actions of $\cP$ and $\cT$ are given in (\ref{E4}), where
the gamma matrices are given in (\ref{E3}). Equation (\ref{ee21}) is form
invariant under the combined operations $\cP$ and $\cT$ because the functions
$\psi(t,{\bf x})$ and $\cPT\psi(t,{\bf x})=\gamma^0(i\gamma^1\gamma^3)\psi^*(-t,
-{\bf x})$ satisfy the same equation. For the free Dirac equation, this is true
for $\cP\psi=\gamma^0\psi(t,-{\bf x})$ and $\cT\psi=i\gamma^1\gamma^3\psi^*(-t,
{\bf x})$ individually. By setting ${\bf x}\to-{\bf x}$ in (\ref{ee21}), it
becomes
$$(i\gamma^0\partial_0-i\gamma^i\partial_i-m)\psi(t,-{\bf x})=0,$$ 
where $i=1,2,3$ denote the spatial components. Multiplying this result from the
left with $\gamma^0$ and using the anticommutation relations $\{\gamma^\mu,
\gamma^\nu\}=2g^{\mu\nu}$, with $g^{\mu\nu}={\rm diag}(1,-1,-1,-1)$ results in 
$$(i\slashed\partial-m)\gamma^0\psi(t,-{\bf x})=0.$$
On the other hand, taking the complex conjugate of (\ref{ee21}) and replacing
$t\to-t$ gives 
$$\big[-i\big(-\gamma^0\partial_0+\gamma^1\partial_1-\gamma^2\partial_2+\gamma^3
\partial_3)-m\big]\psi^*(-t,{\bf x})=0$$
because $\big(\gamma^2\big)^*=-\gamma^2$. Multiplying this equation from the
left by $i\gamma^1\gamma^3$ and using the anticommutation relations for the
gamma matrices then gives
$$(i\slashed\partial-m)i\gamma^1\gamma^3\psi^*(-t,{\bf x})=0.$$
The form invariance of the equation satisfied by $\cPT\psi(t,{\bf x})$ then
follows.

Next we include an axial non-Hermitian bilinear term into the Lagrangian
density, ${\cal L}={\cal L}_0+{\cal L}_{\rm int}$, where ${\cal L}_{\rm int}=-g
\bar\psi\gamma^5\psi$. The corresponding Euler-Lagrange equation 
\begin{equation}
(i\slashed\partial-m-g\gamma^5)\psi(t,{\bf x})=0,
\label{eee21}
\end{equation}
superficially resembles the 1+1-dimensional case. However, here, while parity
transforms this equation into
\begin{equation}
(i\slashed\partial-m+g\gamma^5)\gamma^0\psi(t,-{\bf x})=0, 
\label{ee22}
\end{equation}
time reversal transforms it into
\begin{equation}
(i\slashed\partial-m-g\gamma^5)i\gamma^1\gamma^3\psi^*(t,-{\bf x})=0.
\label{ee23}
\end{equation}
Note the minus sign before the last term in (\ref{ee23}): While parity flips the
sign of the axial term, time reversal in 3+1 dimensions does not. Parity is odd,
but time reversal is {\it even} in 3+1 dimensions. So the combination of $\cPT$
does not lead to a form-invariant Dirac equation. The axial term by itself is
anti-$\cPT$ symmetric. This differs from the 1+1-dimensional case [see
(\ref{E9}) and (\ref{E10})].

The dispersion relation that one obtains from (\ref{eee21}) is formally the same
as in the 1+1-dimensional case; assuming plane-wave solutions of the form $\psi
=e^{-ip^\mu x_\mu}$ and multiplying (\ref{eee21}) by $(\slashed p+m+g\gamma^5)$,
we arrive at the same spectral relation as in 1+1 dimensions, 
$$p^2=m^2-g^2,$$
which is positive only when $m^2\geq g^2$ and is complex in the chiral limit
$m\to0$. 

As before, the form invariance of the Dirac equation unter $\cPT$ imples that
$H(\cPT\psi)=\cPT (H\psi)$, where $H$ is the Dirac Hamiltonian identified
through $i\partial_t\psi=H\psi$. Thus, we can ascertain the properties of
various interaction terms by testing them with this commutation relation. For
(\ref{eee21}) the associated Hamiltonian is
$$H=\alpha\cdot({-i\bf\nabla})+\beta m+\beta g\gamma^5.$$
Let us check the symmetry of the axial interaction term $H_{\rm int}=g\gamma^0
\gamma^5$ under $\cP$ and $\cT$. Using (\ref{E4}), we evaluate $\cPT\psi(t,{\bf
x})=\gamma^0 i\gamma^1\gamma^3\psi^*(-t,-{\bf x})$ and apply $H_{\rm int}$:
\begin{eqnarray}
H_{\rm int}\big(\cPT\psi\big)&=& g\gamma^0\gamma^5\gamma^0 i\gamma^1\gamma^3
\psi^*=-\gamma^0 i\gamma^1\gamma^3 g\gamma^0\gamma^5\psi^*\nonumber\\
&=& -PTH_{\rm int}\psi^*=-\cPT H_{\rm int}^*\psi \nonumber \\
&=& -\cPT\big(H_{\rm int}\psi\big).
\label{e25}
\end{eqnarray}
$H_{\rm int}$ {\it anti}-commutes with $\cPT$, confirming that this term is
not $\cPT$ symmetric. It thus explains the complex nature of the dispersion
relation in the chiral limit. By contrast, if $H_{\rm int}$ is imaginary, that
is $H_{\rm int}=ig\gamma^0\gamma^5$, we have a $\cPT$-symmetric Hamiltonian,
which is also Hermitian, and does have a real spectrum for all $g$, $p^2=g^2$ in
the chiral limit.

Once again, to clarify this point, we turn to an explicit matrix representation.
Then $H_{\rm int}$ becomes
\begin{equation}
H_{\rm int}=g\begin{pmatrix} 0 & 0 & 1 & 0 \\ 0 & 0 & 0 & 1 \\ -1 & 0 & 0 & 0 \\
0 & -1 & 0 & 0\end{pmatrix},
\label{e26}
\end{equation}
which is not Hermitian. By comparison, a general four-dimensional
$\cPT$-symmetric fermionic Hamiltonian that is invariant under $\cPT$ and also
selfadjoint under the $\cPT$ inner product has a matrix form
\cite{r12,r13,abspk},
\begin{equation}
H=\begin{pmatrix} a_0 & 0 & -C_- & -B_-\\ 0 & a_0 & - B_+ & C_+\\
C_+ & B_- & -a_0 & 0\\ B_+ & -C_- & 0 & -a_0\end{pmatrix},
\label{e27}
\end{equation}
where $B_{\pm}=b_1\pm ib_2$ and $C_{\pm}=b_3\pm ib_0$. The parameters $a_0$,
$b_0$, $b_1$, $b_2$, and $b_3$ are real. This matrix has twofold degenerate
real eigenvalues
\begin{equation}
E_\pm=\pm\sqrt{a_0^2-b_0^2-b_1^2-b_2^2-b_3^2},
\label{e28}
\end{equation}
for $a_0^2\ge\sum_{i=0}^3 b_i^2$ \cite{footnote}. Equation (\ref{e26}) is not a
special case of (\ref{e27}), so it is does not represent a $\cPT$-symmetric
fermionic Hamiltonian.

Evidently, the symmetry properties of the axial term $-g\gamma^5\psi$ in the
Dirac equation in 1+1 dimensions differ from those in 3+1 dimensions. The Dirac
equation is form invariant in 1+1 dimensions under $\cPT$, but not in 3+1
dimensions. This corresponds to a relativistic $\cPT$-symmetric
quantum-mechanical Hamiltonian in 1+1 dimensions, but not in in 3+1 dimensions.
This difference is caused by the different effect of time reversal in 1+1 and
3+1 dimensions. The spectrum obtained in both cases is formally the same, so we
conclude that the $\cPT$ symmetry is always broken in 1+1 dimensions when $m\to
0$. However, in 3+1 dimensions the Hamiltonian is anti-$\cPT$ symmetric in the
chiral limit, which explains the complex nature of the spectrum when $m\to0$.

Interestingly, if we include the conventional mass term $m\gamma^0$, (\ref{e26})
becomes
\begin{equation}
H_{\rm int}=\left(\begin{array}{cccc} m & 0 & g & 0\\ 0 & m & 0 & g\\ -g & 0 & 
-m & 0\\ 0 & -g & 0 & -m\\ \end{array} \right),
\label{100200}
\end{equation}
which is neither Hermitian nor $\cPT$-symmetric. However, $H_{\rm int}$ is {\it
pseudo}-Hermitian in the sense of \cite{GWM} because $H_{\rm int}^\dag=\gamma^0
H_{\rm int}(\gamma^0)^{-1}$. Hence, this Hamiltonian can be used to describe
pseudo-Hermitian fermions.

We can construct fermionic creation and annihilation operators which are
quadratically nilpotent, and investigate their anticommutation relations. First,
we note that the eigenvalues of (\ref{100200}) are
$$E_\pm=\pm\omega=\pm\sqrt{m^2-g^2},$$
with corresponding eigenvectors
\begin{widetext}
$$\big|E_-^{(1)}\big\rangle=\frac{1}{\sqrt{2w}}\left(\begin{array}{cccc}0\\
-\sqrt{m+w}(m-w)/g\\ 0\\ \sqrt{m+w}\\ \end{array}\right)\qquad\quad
\big|E_-^{(2)}\big\rangle=\frac{1}{\sqrt{2w}}\left(\begin{array}{cccc}-
\sqrt{m+w}(m-w)/g\\ 0\\ \sqrt{m+w}\\ 0\\ \end{array}\right),$$
$$\big|E_+^{(1)}\big\rangle=\frac{1}{\sqrt{2w}}\left(\begin{array}{cccc} 0\\
-\sqrt{m-w}(m+w)/g\\ 0\\ \sqrt{m-w}\\ \end{array}\right) \qquad\quad
\big|E_+^{(2)}\big\rangle=\frac{1}{\sqrt{2w}} \left(\begin{array}{cccc}
-\sqrt{m-w}(m+w)/g\\ 0\\ \sqrt{m - w}\\ 0\\ \end{array} \right).$$
\end{widetext}
The spectrum is twofold degenerate and is real if $g^2\leq m^2$. This degeneracy
is the analog of the phenomenon of Kramer's theorem in conventional Hermitian
quantum mechanics, where the Hamiltonian is invariant under odd time reversal,
as is the case with (\ref{100200}).

We introduce the annihilation operator for the Hamiltonian (\ref{100200}) as
$$\eta=\frac{1}{2w} \left(\begin{array}{cccc} g & 0 & m-w & 0\\ 0 & g & 0 & m-w
\\ -m - w & 0 & -g & 0\\ 0 & -m - w & 0 & -g\\ \end{array} \right),$$
which is nilpotent ($\eta^2=0$) as required. We verify that
$$\eta\big|E_-^{(1)}\big\rangle=\eta\big|E_{-}^{(2)}\big\rangle=0,$$
$$\eta\big|E_+^{(1)}\big\rangle=\big|E_-^{(1)}\big\rangle,\qquad\quad
\eta\big|E_+^{(2)}\big\rangle=\big|E_-^{(2)}\big\rangle.$$
The creation operator reads 
$$\eta^\prime=\frac{1}{2w}\left(\begin{array}{cccc} g & 0 & m+w & 0\\ 0 & g & 0
& m+w\\ -m+w & 0 & -g & 0\\ 0 & -m+w & 0 & -g\\ \end{array}\right).$$
One can now establish the anticommutation relations
$$\{N,\eta\}=-\eta\qquad\quad \{N,\eta^{\prime}\}=-\eta^{\prime},$$
where $N$ is the number operator, $N=\eta^{\prime}\eta$, as well as the peculiar
anticommutation relation $\eta\eta^\prime+\eta^\prime\eta=-\mathbbm{1}$. The
minus sign indicates that the number operator gives the negative of the state
occupation number. For further illustrations of this in the context of $\cPT$
symmetry see Refs.~\cite{r14,R19}.

Finally, we comment that in terms of the number operator $N$, we can write the
four-dimensional pseudo-Hermitian fermionic Hamiltonian in (\ref{100200}) in the
form of a free (bosonic) harmonic oscillator as
$$H=\Delta\omega(-N)+\omega_-\mathbb{1},$$
where $\Delta\omega=\omega_+ -\omega_-$ and $\mathbb{1}$ is the four-dimensional
identity matrix.

\subsection{Other matrix-type two-body (four-point) $\cPT$- and
anti-$\cPT$-symmetric interactions and the resulting $\cPT$-symmetric
Hamiltonians}\label{s3b}
Having determined that an axial non-Hermitian interaction Lagrangian density of
the form $-g\bar\psi\gamma^5\psi$ in 3+1 dimensions does not give rise to a
Dirac equation that is form invariant with respect to $\cPT$, we seek other
types of interactions that are $\cPT$ symmetric but non-Hermitian. Usually, the
standard method of analyzing two-body (four-point) interactions involves
constructing the 16 independent bilinears from the 16 $4\times4$ independent
matrices and considering the Lagrangian density associated with each of these.
The standard Hermitian combinations are (1) $\bar\psi\psi$, (2) $\bar\psi
\gamma^\mu\psi$, (3) $\bar\psi\sigma^{\mu\nu}\psi$, (4) $\bar\psi\gamma^5
\gamma^\mu\psi$, and (5) $i\bar\psi\gamma^5\psi$. This Lagrangian-density
approach is suitable for a discussion of symmetries that lead to conserved
currents through Noether's theorem, but the analysis of $\cPT$ symmetry is most
simply done by examining the form-invariance of the appropriate Dirac-like
equation that can be derived using the Euler-Lagrange equations. Since this in
turn tranlates into a commutation relation of the Hamiltonian with $\cPT$, in a
form of {\it reverse engineering}, we only need to identify possible
$\cPT$-symmetric forms of the interaction Hamiltonians. Thus, we consider the
five interaction Hamiltonians below and show that these combinations are all
$\cPT$ symmetric:
\begin{eqnarray}
H_{{\rm int},1} &=& g \gamma^0,\nonumber\\
H_{{\rm int},2} &=& B_\mu \gamma^0 \gamma^\mu,\nonumber\\
H_{{\rm int},3} &=& i T_{\mu\nu}\gamma^0 \sigma^{\mu\nu},\nonumber\\
H_{{\rm int},4} &=& i\tilde B_\mu \gamma^0 \gamma^5\gamma^\mu,\nonumber\\
H_{{\rm int},5} &=& ig_A \gamma^0 \gamma^5,\nonumber
\end{eqnarray}
where $g,\,B_\mu,\,T_{\mu\nu},\,\tilde B_\mu,$ and $g_A$ are taken to be real.

Using the procedure in (\ref{e25}) in which $H_{{\rm int},i}$ is applied to
$\cPT\psi$, we evaluate the commutator of $H_{{\rm int},i}$ and $\cPT$ using
(\ref{E4}), and where necessary, make use of the relation $\gamma^\mu i
\gamma^1\gamma^3=i\gamma^1\gamma^3\gamma_{\mu}^*$. Then
\begin{widetext}
\begin{eqnarray}
H_{{\rm int},1}(\cPT\psi) &=& g\gamma^0\gamma^0i\gamma^1\gamma^3\psi^*
=\gamma^0i\gamma^1\gamma^3 g\gamma^0\psi^*
=PTH_{{\rm int},1}\psi^*=\cPT(H_{{\rm int},1}\psi),\\ 
H_{{\rm int},2}(\cPT\psi) &=& B_\mu\gamma^0\gamma^\mu\gamma^0i\gamma^1
\gamma^3\psi^*=\gamma^0i\gamma^1\gamma^3B_\mu\gamma^0\gamma^{\mu*}\psi^*
=PTH_{{\rm int},2}^*\psi^*=\cPT(H_{{\rm int},2}\psi),\\
H_{{\rm int},3}(\cPT\psi) &=& iT_{\mu\nu}\gamma^0\sigma^{\mu\nu}\gamma^0
i\gamma^1\gamma^3\psi^*
=-\gamma^0i\gamma^1\gamma^3iT_{\mu\nu}\gamma^0\sigma^{\mu\nu*}\psi^*
=-PTi\gamma^0\sigma^{\mu\nu*}T_{\mu\nu}\psi^* =\cPT (H_{{\rm int},3}\psi),\\
H_{{\rm int},4}(\cPT\psi) &=& i\tilde B_\mu\gamma^0\gamma^5\gamma^\mu\gamma^0i
\gamma^1\gamma^3\psi^*
=\gamma^0i\gamma^1\gamma^3(-i)\tilde B_\mu\gamma^0\gamma^5\gamma^{\mu*}\psi^*
= PT H_{{\rm int},4}^*\psi^* = \cPT (H_{{\rm int},4}\psi),\\
H_{{\rm int},5}(\cPT\psi) &=& ig_A \gamma^0\gamma^5\gamma^0i\gamma^1
\gamma^3\psi^* =\gamma^0i\gamma^1\gamma^3(-i)g_A\gamma^0\gamma^5\psi^*
=PT H_{{\rm int},5}^*\psi^*=\cPT(H_{{\rm int},5}\psi).
\end{eqnarray} 
\end{widetext}
We conclude that
$$[\cPT, H_{{\rm int},i}]=0\quad(i=1,\,\cdots,\,5).$$

Thus, the general form of a relativistic quantum-mechanical Dirac equation,
which is form invariant under $\cPT$ transformations, reads
$$(i\slashed\partial-g-B_\mu\gamma^\mu-iT_{\mu\nu}\sigma^{\mu\nu}-i\tilde B_\mu
\gamma^5\gamma^\mu-ig_A\gamma^5)\psi(t,{\bf x})=0.$$
A brief analysis shows that $H_{{\rm int},3}$ and $H_{{\rm int},4}$ are
anti-Hermitian, while $H_{{\rm int},1}$, $H_{{\rm int},2}$, and $H_{{\rm int}
,5}$ are Hermitian. So we have identified two types of terms that give rise to
non-Hermitian but $\cPT$-symmetric Hamiltonians. We consider each of these in
turn.

\subsubsection{$H_{{\rm int},3}=iT_{\mu\nu}\gamma^0\sigma^{\mu\nu}$}
To understand the structure of $H_{{\rm int},3}$ we write it in matrix form:
\begin{equation}
H_{{\rm int},3}=\begin{pmatrix} iq_4 & -q_5+iq_6 & - q_3 & -q_1+iq_2\\
q_5 +iq_6 & -iq_4 & -q_1-iq_2 & q_3\\ q_3 & q_1-iq_2 & -iq_4 & q_5-iq_6\\
q_1+iq_2 & -q_3 & -q_5-iq_6 & iq_4\end{pmatrix},
\label{e29}
\end{equation}
where the coefficients $q_i$, $i=1,\dots, 6$, are abbreviations for combinations
of the $T_{\mu\nu}$, 
\begin{eqnarray}
&& q_1=T_{01}-T_{10},\quad q_2=T_{02}-T_{20},\quad q_3=T_{03}-T_{30},\nonumber\\
&& q_4=T_{12}-T_{21},\quad q_5=T_{13}-T_{31},\quad q_6=T_{23}-T_{32}.\nonumber
\end{eqnarray}
The eigenvalues of (\ref{e29}) are
\begin{eqnarray}
&&\pm\big\{-Q^2\pm2\big[\big(q_1^2+q_2^2\big)q_4^2+\big(q_1^2+q_3^2\big)q_5^2
+\big(q_2^2+q_3^2\big)q_6^2\nonumber\\
&&+2q_2q_3q_4q_5+2q_1q_2q_5q_6-2q_1q_3q_4q_6\big]^{1/2}\big\}^{1/2},
\nonumber
\end{eqnarray}
where $Q^2=\sum_{i=1}^6 q_i^2$. Thus, the eigenvalues are complex and the $\cPT$
symmetry is broken. Including a finite mass term $m\gamma^0$ in general does not
change this result. The eigenvalues of $H_{{\rm int},3}+m\gamma^0$ are modified
to read
\begin{eqnarray}
&&\pm\big\{m^2-Q^2\pm 2\big[\big(q_1^2+q_2^2-m^2\big)q_4^2+\big(q_1^2+q_3^2-m^2
\big)q_5^2\nonumber\\
&& ~~+\big(q_2^2+q_3^2-m^2\big)q_6^2+2q_2q_3q_4q_5+2q_1q_2q_5q_6\nonumber\\
&& ~~-2q_1q_3q_4q_6\big]^{1/2}\big\}^{1/2}.\nonumber
\end{eqnarray}
As we have already argued, only if the spectrum is twofold degenerate, can the eigenvalues be real \cite{footnote}.

\begin{figure*}
\centering
\includegraphics[scale=0.5]{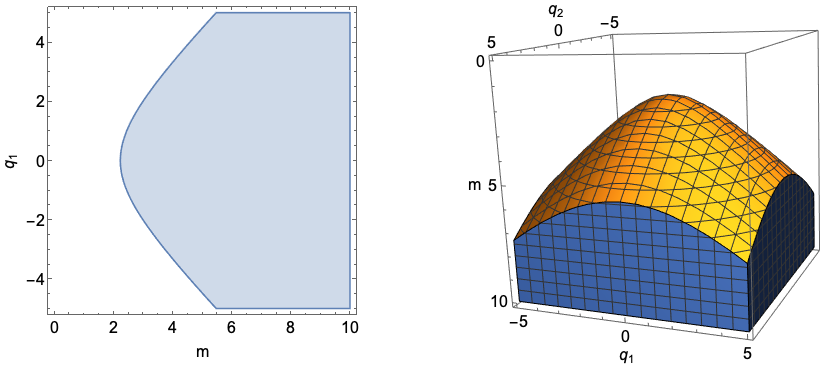}
\caption{Parametric regions of unbroken $\cPT$ symmetry (shaded regions) for the
Hamiltonian $H_{\rm int,3}+m\gamma^0$, where $q_4=q_5=q_6=0$. Left panel: in the
$(m,q_1)$ plane, $q_2=1$ and $q_3=2$. Right panel: $q_3=2$.}
\label{f17}
\end{figure*}

If we compare (\ref{e29}) with (\ref{e27}), we see that both have a
quaternionic structure. However, in addition to being $\cPT$ symmetric,
(\ref{e27}) fulfills the additional condition that this Hamiltonian is
selfadjoint with regard to the $\cPT$ inner product according to
\cite{r12}. This means that, in addition, $H_{{\rm int},3}$ should fulfill
the condition $H_{{\rm int},3}^{\cPT} = P H_{{\rm int},3}^\dagger P =
H_{{\rm int},3} $. If we construct $H_{{\rm int},3}^{\cPT}$, we find that
$$q_4=q_5=q_6=0,$$
for this condition to hold. The eigenvalues are twofold degenerate and if a mass
term is included, they are
$$E_\pm=\pm\sqrt{m^2-q_1^2-q_2^2-q_3^2},$$
which is real provided that $m^2\ge q_1^2+q_2^2+q_3^2$. Thus, $\cPT$ symmetry is
broken in the chiral limit. The regions of unbroken $\cPT$ symmetry for the
Hamiltonian $H_{\rm int,3}+m\gamma^0$ for some specific parameters are shown in
Fig.~\ref{f17}.

\subsubsection{$H_{{\rm int},4}=i\tilde B_\mu\gamma^0\gamma^5\gamma^\mu$}
We now consider the equation of motion resulting from the non-Hermitian
$\cPT$-symmetric Hamiltonian $H_{{\rm int},4}$ (as well as its corresponding
Lagrangian $\cL_{{\rm int},4}$),
$$(i\slashed\partial-i\gamma^5\slashed{\tilde B})\psi=0.$$
The spectrum associated with this equation can be obtained by calculating the
poles of the associated Green function in momentum space, which satisfies 
$$(\slashed p-i\gamma^5\slashed{\tilde B})S(p)=1.$$
Rationalizing this expression for $S(p)$, we identify the dispersion relation as
$$(p^2-{\tilde B}^2)^2+4(p\cdot\tilde B)^2=0.$$
This has no real solutions for all $p_0$. Thus, again we find that the $\cPT$
symmetry of the Hamiltonian is broken. We also notice that an
anti-$\cPT$-symmetric but Hermitian Hamiltonian would give a real spectrum with
dispersion relation $(p^2-\tilde B^2)^2-4(p\cdot\tilde B)^2=0$.

Note that the matrix form of the Hamiltonian $H_{{\rm int},4}$, with components 
$\tilde B_\mu=(\tilde B_0,\tilde B_1,\tilde B_2,\tilde B_3)$ is 
\begin{eqnarray}
&& H_{{\rm int},4}=\nonumber\\
&& \left(\begin{array}{cccc} -i\tilde B_3 & -i\tilde B_1-\tilde B_2 & -i\tilde
B_0&0\\ \tilde B_2-i\tilde B_1 & i\tilde B_3& 0 &-i\tilde B_0\\ -i\tilde B_0 &
0 & -i\tilde B_3 &-i\tilde B_1-\tilde B_2\\ 0 &-i\tilde B_0 &\tilde B_2-i
\tilde B_1 &i\tilde B_3\\ \end{array}\right),\nonumber
\end{eqnarray}
which has complex eigenvalues for all $\tilde B_\mu$ real,
\begin{eqnarray}
E_{1,2} &=& i\tilde B_0\pm i\sqrt{\tilde B_1^2+\tilde B_2^2+\tilde B_3^2},
\nonumber\\
E_{3,4} &=& -i\tilde B_0\pm i\sqrt{\tilde B_1^2+\tilde B_2^2+\tilde B_3^2}.
\nonumber
\end{eqnarray}

If, as in Subsec.~IIIB1, we demand that the Hamiltonian $H_{{\rm int},4}$
satisfies the selfadjointness condition according to \cite{r12,r13,R19}, that
is, $H_{{\rm int},4}^{\cPT}=PH_{{\rm int},4}^\dag P=H_{{\rm int},4}$, we
calculate that $\tilde B_0\neq0$ and $\tilde B_1=\tilde B_2=\tilde B_3=0$. The
resulting twofold degenerate energies are
\begin{equation}
E_\pm=\pm\sqrt{m^2-\tilde B_0^2},
\label{e30}
\end{equation}
where we have included a mass term. This implies a real spectrum for $m^2\ge
\tilde B_0^2$. Once again, in the chiral limit the $\cPT$ symmetry is broken.
 
\section{Main conclusions and outlook}\label{s4}
Our focus in this paper has been on investigating non-Hermitian $\cPT$-symmetric
extensions to fermionic systems in 1+1 and 3+1 dimensions. The main findings
are the following:

a) Usually, we explore the symmetries of a field theory by examining the
Lagrangian density. However, the properties associated with $\cPT$ symmetry are
more easily found by forming the Euler-Lagrange equations and demanding
form-invariance of the relativistic equation of motion with respect to $\cPT$.
This is equivalent to constructing the quantum-mechanical relativistic
Hamiltonian and investigating its commutation relation with $\cPT$. 

b) For a pure axial interaction the symmetry properties in 1+1 dimensions
differ from those in 3+1 dimensions even though the formal structure of the
energy relation is unchanged. This can be traced back to the different
transformation properties of time reversal in 1+1 and 3+1 dimensions and is
ultimately due to the fact that $\cT^2=-\one$ in 3+1 dimensions.

c) In 1+1 dimensions including a complex $\cPT$-symmetric position-dependent
potential in both scalar- and vector-coupling schemes and combinations thereof 
can result in real and discrete eigenvalues, when searching for plane wave
solutions. For appropriately chosen combinations of scalar and vector couplings,
a Schr\"odinger-like equation can be found and the spectrum can be determined
numerically. The analogous classical systems give information about the nature
of the spectrum. They display closed contours when the eigenvalues are real and
discrete and they are periodic and open if there is a real band structure. If
the eigenvalues are complex, the paths are open and nonperiodic.

d) In 3+1 dimensions only two possible Lorentz-invariant two-body combinations
are $\cPT$ symmetric and not Hermitian. These, however, give rise to a complex
spectrum in the chiral limit. Including a mass term can result in a real
spectrum. In addition, further constraints are placed on the parameters if the
condition of selfadjointness with respect to the $\cPT$ inner product is placed
on the Hamiltonian. This does not change the conclusion. 

It remains an open question as to whether including non-Hermitian
$\cPT$-symmetric terms can play a role in physical fermionic systems, for
example, affecting chiral symmetry restoration within the Nambu-Jona-Lasinio or
Thirring models, or in weak interactions. 

\acknowledgments
CMB thanks the Alexander von Humboldt Foundation for financial support and the
Heidelberg Graduate School for Physics for its kind hospitality.


\begin{thebibliography}{99}
\bibitem{R20} In this paper most of the Hamiltonians that are considered are
bilinear in the fermionic field. One result of this paper is that it is not
possible to construct a $\cPT$-symmetric bilinear fermionic Hamiltonian that has
a real spectrum in the chiral limit. However, we firmly believe that a
Hamiltonian having higher-power interactions will achieve this goal. A simple
bosonic model that illustrates this point is the Hamiltonian $H=p^2+\big(m^2-
\mu^2\big)x^2-gx^4$. In the quadratic case $g=0$ we see that the spectrum
becomes complex in the $m\to0$ limit. However, if $g\neq0$ it is shown in
C.~M.~Bender, D.~C.~Brody, J.-H.~Chen, H.~F.~Jones, K.~A.~Milton, and
M.~C.~Ogilvie, Phys.~Rev.~D {\bf 74}, 025016 (2006) that $H$ is equivalent to
the Hermitian Hamiltonian
$\tilde H=p^2-2\hbar x\sqrt{g}+4g\big[x^2-\big(m^2-\mu^2\big)/(4g)\big]^2$,
which has a spectrum that is real and bounded below for {\it all}
values of $m$ and $\mu$.
\bibitem{footnote} When $H$ and $\cPT$ commute, in general, in order to have a real spectrum, we have two possibilities. One option is $\psi$ to be $\cPT$ symmetric, i.e., it has the symmetry of the Hamiltonian. The other option is that $\psi$ and $\cPT \psi$ be degenerate eigenvectors of $H$. For $\cT^2= - \one$ the first option is ruled out since, if we require $\cPT\psi=\psi$, then $\psi$ must vanish; So the degeneracy of the Hamiltonian is a necessary condition for the spectrum to be real.
\end{thebibliography}
\end{document}